# Indigenous Astronomies and Progress in Modern Astronomy

**Clive Ruggles**[1]

*School of Archaeology and Ancient History, University of Leicester*
*University Road, LEICESTER  LE1 7RH, U.K.*
*E-mail:* `rug@le.ac.uk`

From an anthropological point of view, the whole concept of a 'path of progress' in astronomical discovery is anathema, since it implicitly downgrades other cultural perspectives, such as the many 'indigenous cosmologies' that still exist in the modern world. By doing so, one risks provoking those who hold them and—as is most obvious in places such as Hawaii where the two 'world-views' come into direct contact—creating avoidable resistance to that very progress. The problem is complicated by the existence of 'fringe' and 'new-age' views that are increasingly confused with, and even passed off as, indigenous perceptions.

In a modern world where widespread public perceptions include many that are unscientific in the broadest sense of the term, I shall argue that there are actually a range of positive benefits for progress in scientific astronomy to be derived from the mutual awareness and comprehension of 'genuine' cultural world-views whose goals—in common with those of modern science—are to make sense of the cosmos within which people live. While two-way education is clearly a prerequisite, I shall argue that the necessary level of reconciliation can only be achieved through more fundamental attempts by modern astronomers to understand, and ultimately to respect, both the non-Western frameworks of thought that give rise to other cultural perspectives and the heritage associated with them. One of the most obvious potential benefits could derive from common attitudes towards the natural heritage of astronomy, namely dark skies.

---

[1] Speaker

# 1. Introduction

How is the progress of modern astronomical research influenced by wider perceptions of astronomy, and of science in general? An obvious answer is that public perceptions affect political expediency, and consequently influence levels of support (both governmental and private) and in particular funding priorities. An awareness of such issues is clear in the IAU's new strategic initiative 'Astronomy for the developing world' [1], one of whose aims is to help enable developing nations to contribute to cutting-edge scientific research. The IAU's Strategic Plan rightly emphasises the inspirational role of astronomy in facilitating education and capacity building as well as furthering sustainable development, and rightly stresses the importance of education and outreach. It is also true (as the Strategic Plan states) that astronomy "embodies a unique combination of science, technology and culture [and] continues to play an important role in modern society". However, in the context of many peoples of the world, and particularly within developing nations, the 'astronomy' that continues to play an important cultural role—in the sense of perceptions of the sky and its importance for the functioning of society—is not modern scientific or (for want of a better term) 'Western' astronomy. In such contexts, there are innate hurdles on the path towards even the acceptance, let alone better comprehension and contribution to the advancement, of modern scientific astronomy. One of the prerequisites is that 'Western' astronomers attempting to communicate 'their' astronomy must acknowledge and respect the importance of these indigenous astronomies; and greater respect can only come from a greater degree of comprehension.

This paper concerns indigenous perceptions of astronomy and attempts to address, in broad terms, how best to establish productive communication, mutual respect and understanding across the barrier or 'divide' that exists between the mindsets of Western scientists and traditional peoples in many developing countries regarding astronomy. I shall go on to argue that, having done this, we can begin to appreciate a range of potential benefits for progress in scientific astronomy that can be derived from the mutual awareness and comprehension of 'genuine' cultural world-views whose goals—in common with those of modern science—are to make sense of the cosmos within which people live, even though they may be embedded within cultural practices that are completely different.

## 2. Approaching the divide

### 2.1 Descending from the pedestal

To 'Western' sensibilities, the traditional calendar of the Mursi—a small group of pastoralists and cattle herders still living in almost complete isolation in the Omo valley of south-western Ethiopia until the 1960s [2,3]—seems haphazard in the extreme, perhaps almost laughable. When asked, everyone including children can recite a chant counting off the months and relating the seasonal activities associated with each month, such as sorghum planting or the coming of the big rains. However, in practice nobody actually knows for certain what month it is at any given time, believing this to be the preserve of 'experts' who, when one tries, can never quite be run to ground. In reality there is always disagreement. This inherent disagreement is not removed at the first appearance of the waxing crescent moon, marking the beginning of a new month, since each school of opinion simply increments their previous estimate. But nor it is resolved when, for example, the big rains come, since everyone knows that the big rains are sometimes early or late: this seasonal event (along with many others) merely serves to shift the balance of opinion. The issue of intercalation—the need to insert an additional month or miss out a whole month from time to time in order to keep the lunar month count in step with the seasons—never arises: it is simply irrelevant, since continual observations of phenomena of nature, such as the flowering of plants or the appearance of birds, lead to continual adjustments.

Preposterous as this 'institutionalized disagreement' may seem as a calendrical regulator from a scientific point of view, from anthropological perspective the Mursi calendar can be recognized as not only self-consistent but completely fit for purpose in the cultural context within which it operates. It is also worthy of our respect as an elegant solution to a set of specific problems of managing time and organizing subsistence activities.

But the Mursi calendar also has a much broader significance. It overturns nearly every assumption that is commonly made (or implicitly assumed) by historians of astronomy, as well as many archaeoastronomers, about the way in which early 'pre-scientific' calendars inevitably developed (e.g. [4]). The tendency is to try to define a 'path of progress' starting with simple counts of lunar synodic cycles, proceeding to luni-solar calendars that adjust the month count so as keep track of the seasonal year (for example by observing stellar heliacal events in order to determine whether to add an intercalary month), thence to the development of solar calendars based upon the changing position of sunrise or sunset along the horizon from a fixed observing place, and finally to the development of formulaic procedures that do not require continual recourse to actual observations.

Suppose, for instance, that an ethnologist working with an indigenous community were to discover the existence of specialists who observe the changing position of sunrise along the horizon over the year. Then, following the 'path of progress' paradigm, this would surely be taken to imply the existence of a solar calendar. However, the Mursi show this not to be the case. Among the Mursi there are indeed 'specialists' (really hobbyists) who track the sunrise along the mountainous horizon beyond the River Omo, watching from their favourite rock or tree. They can, for example, identify its upstream and downstream 'houses' (where it appears to rise at the same position for several weeks around the solstices) and can typically identify the half of the month in which the sun enters and leaves each of its houses. One can almost imagine well-meaning commentators, on hearing this, proclaiming that the Mursi did in fact have a solar calendar: that they had, in fact, advanced a rung up the ladder towards modern science.

It might seem surprising, then, that the Mursi 'sun watchers' are accorded no special social status and tend to share this interest with, for example, that of divining the future using goat entrails. Surprising, that is, until one realizes that, as the Mursi see it, the sun is no more reliable a seasonal marker than many other phenomena of nature such as the flowering of plants and the appearance of birds. It can sometimes be early or late entering or leaving one of its houses (and indeed, it is evident from the Western perspective that it *cannot* always enter or leave in the same half of a month, since the lunar phases are not correlated with the solar year).

The example of the Mursi calendar not only shows the dangers of extending the 'path of progress' paradigm outside the confines of modern scientific astronomy. It also draws attention to the fundamental difference of approach that exists between historians of astronomy—who inevitably seek to identify paths of progress towards modern astronomy—and anthropologists and archaeologists, who strive to understand human achievements in their own terms and to emphasize cultural diversity (cf. [5]). This difference in emphasis has been reflected in numerous debates between 'cultural astronomers' (ethnoastronomers and archaeoastronomers) and historians of astronomy: for example, a 'path of progress' approach can be implicit (or perceived as implicit) in terms commonly used in archaeoastronomy such as 'ancient observatory' and even 'ancient astronomy' itself [6]. It also extends to heritage issues (which are often intimately bound up with contemporary cultural issues): the existence of the 'oldest observatory' in some region of the world is often claimed in support of misguided concepts of national identity and pride [7].

The more insidious problem with extending any sort of 'path of progress' approach outside the framework of modern 'Western' science—whether applied to past cultures or to contemporary indigenous ones—is that it implicitly places our own culture on a pedestal, measuring other cultures against the milestones of our own achievement. In the case of past cultures, this relates to ethnocentrism, the tendency (all too evident among early archaeoastronomers) to project our own ways of comprehending things onto the people we are interested in [8], and has often been driven, ironically, by a desire to highlight the achievements of the past cultures by demonstrating their level of sophistication (again, as judged in our own terms).

If our aim is to understand something of indigenous perceptions of the cosmos, or indeed of our own science, then this must clearly involve establishing a dialogue. When one is dealing face-to-face with people from indigenous communities, any attempt (explicit or implicit) to place one's own culture on a pedestal will almost inevitably be taken as condescending and patronising. It follows that if we are to take the first steps towards understanding other cultural perspectives relating to the natural world and the sky, then we must genuinely abandon any thought (or any implicit assumption) that our own scientific viewpoint is superior, rather than just different.

## 2.2 Avoiding cultural relativism

The problem with this approach is that it might seem to imply a sort of cultural relativism that acknowledges any point of view, or any evidence (or imagined evidence) interpreted in any way, as being as equally valid as the scientific consensus.

Imagine yourself at a 'cultural astronomy' conference listening to a paper on, let us say, indigenous shamans and how they manipulate the sky and weather. Speaker X is a member of the indigenous community in question but impresses the (Western) audience with his detailed and scholarly study and analysis of a rain-making ritual. Only then does he make a passing remark that makes it suddenly obvious that the speaker himself believes that the shaman really can create rain. A silent shock wave ripples through the audience, who at once become wary of speaker X who has instantly, through this one remark, annihilated his scientific credentials.

This fictitious scenario is in fact very close to two or three situations that the author has encountered at first hand in 'cultural astronomy' conferences, and illustrates not only the existence of the divide (of which speaker X has suddenly placed himself on the other side) but also the difficulty of maintaining 'respect' among 'Western' academics when one reveals oneself as not necessarily sharing the same 'rationality'.

The basic issue here is that 'Western' academic discourse is founded upon a set of general principles that, without needing to enter any detailed debates about scientific method, would surely be universally characterised as 'scientific' in the broadest sense: attempting to assess fairly all the available evidence, always being open-minded towards new evidence, actively seeking new evidence, and so on. If someone holds, or appears to hold, beliefs that contradict well-established scientific evidence, then this surely calls into question the credibility of their wider arguments.

Of course, while the discourse is 'among ourselves', there is no problem. The Western anthropologist does not have to accept the tenets of another world-view while striving to understand pertinent aspects of it and to discuss his or her interpretations with other Western academics. (This may seem a trivial point, but it is surprisingly clouded in the literature.) Field anthropologists can still operate entirely within the Western academic framework of interpretation when they communicate with 'informants' beyond the confines of that framework, following established principles and methods when doing so (see §3.2 below). Clearly, we do not have to embrace cultural relativism to the extent of abandoning our own principles of scientific endeavour and academic scholarship in order to develop Western models of non-Western mindsets.

But how should one react to this non-Western scholar seeking an academic dialogue 'across the divide'? How can one genuinely abandon a feeling of superiority, this being (as already argued) an essential prerequisite for meaningful discourse, without abandoning the assumption that our own scientific world-view is superior in that the scientific method (interpreted broadly) is 'correct' and superior to non-rational methods?

The answer, certainly, comes from mutual respect. This may be achieved more easily on the 'Western' side by bearing in mind that other world-views cannot be summarily dismissed as 'non-science', because they *do* make sense of the evidence available to them in a logical way. We shall say more about this in the following section. 'Their' rationality may not be ours, but it is a valid rationality nonetheless.

Once this is achieved, then each side should be able to accept that, while they continue to believe that their own rationality is the correct one, the other side has valid reasons for feeling the same about *their* rationality. In this way, symmetry is established: each side becomes, in a sense, the anthropologist studying the other. Yet even now we have arguably achieved no more than two sides each observing the other and discussing their interpretations with others back in their own community—from the Western perspective it moves us no further forward than the earlier generations of Western ethnographers who 'studied' other cultures.

The critical final, and most challenging, step in the process is to establish exchange at the 'meta' level: the exchange of ideas and interpretations between the two communities with their different ways of rationalising those ideas. At this point we enter realms of anthropological debate that generated considerable interest during the 1970s and 1980s (e.g. [9]) and are sustained to this day through organisations such as the Association of Indigenous Anthropologists (www.aaanet.org). Observing that anthropology is itself a product of Western culture, the idea was to encourage non-Western anthropologists to train in Western universities and then return to their own countries to practice as anthropologists there. An early example was the Ethiopian anthropologist Asmarom Legesse, whose 1973 analysis of the Gada system, a complex African social institution among the Borana people of southern Ethiopia and northern Kenya [10], included an account of the Borana calendar that generated a protracted debate among both anthropologists and archaeoastronomers, and itself highlighted the problems of merging concepts of Western and indigenous astronomies in order to achieve a plausible and sustainable interpretation [11]. Within ethnoastronomy, a modern continuation of the 'indigenous anthropologist' approach is being achieved in several non-Western countries through the activities of people of local descent trained as professional astronomers and astrophysicists (e.g. [12]).

In planning the Special Session, the organizers discussed at length the possibility of inviting an indigenous speaker to present aspects of their world-view, but in the end it was acknowledged that this might not be done in terms that the audience could readily understand. In one sense, the organizers were right to recognize that establishing a meaningful level of academic dialogue is non-trivial and probably could only be achieved after an extended process of mutual 'enculturation'.

In order to prevent this seeming an insurmountable obstacle, it can be extremely helpful from the Western perspective to recognise ways in which other world-views are likely to differ fundamentally from one's own. A useful final stage in approaching the divide, in other words, is to recognise that some of the strongest and most basic tenets of the Western world-view are unlikely to form basic tenets of a different one. But can we 'deconstruct' enough of the Western world-view to be able to identify some of these? The answer is yes, and—even better—in order to do this, we can use to Western 'models' of other frameworks of conceptualisation.

### 2.3 Cosmologies: 'ours' and 'theirs'

One of the most basic assumptions we make about the 'world' (universe/cosmos) is that it exists as an objective reality that we, as intelligent entities, can observe, describe and attempt to understand. We also tend to make a more subtle range of assumptions, for example that the universe is subject to 'rules' or constraints ('physical laws') that have some degree of universal applicability, rather than just being chaotic. It is such assumptions that underlie the whole enterprise of observational/experimental and theoretical science.

Yet anthropology shows clearly that non-Western viewpoints do not generally share such foundations. In typical indigenous perceptions, the 'natural world' does not exist as a separate objective reality in which the individual can be conceived as a passive observer, but as something that one's consciousness is submerged within, and we must develop concepts such as 'lifeworld', 'dwelling' and 'being-in-the-world' to describe this [13]. The individual becomes an 'agent in an environment' rather than a passive observer. (To reinforce the point made earlier about cultural relativism: we do not ourselves have to believe that the natural world does not exist independently of our minds in order to appreciate, and try to understand, the perceptions of those who do.)

This is not to deny that indigenous peoples attempt to make sense of their lifeworld by structuring it—far from it. It is just that they do not use Cartesian dualities (mind/body, subject/object) or Linnaean categorisations. Instead, they establish interconnections and create cultural taxonomies on the basis of social agreement and use [14], using 'dialectic' rather than 'rationalistic' logic [15]. Neither will they generally recognise what we would see as a fundamental distinction between 'empirical reality' and mental constructs (spirit worlds etc).

As twentieth-century anthropology has revealed, even the most technologically unsophisticated human communities have generally developed rich and complex systems that assign meaning and significance to what they perceive as existing in the world [13]. Furthermore these 'indigenous cosmologies' are not passive but prescribe and constrain social action, resulting in preferences for food, dress, plants and animals, marriage partners, and so on. They can also provide a basis for action, for example stipulating the places and times where certain activities should takes place, routes to be taken through the landscape, etc. All of this might be perceived as 'keeping in harmony with the cosmos', but we can often see very direct social consequences or 'uses' of cosmological knowledge, from providing a useful prescription for seasonal activities among a small group that ensures successful subsistence and ultimately survival, through to sustaining structures of social or political power in more complex societies.

In the sense that they represent the attempts of a human society to make sense of the cosmos or lifeworld, albeit within the framework of an alternative rationality, indigenous cosmologies clearly represent 'science' in its broadest sense. This is consistent with the statement, made in the context of UNESCO global heritage, that "In a global context, 'science' may be defined broadly as 'knowledge or systems of knowledge, including traditional knowledge'" [16].

### 3. Bridging the divide

#### 3.1 The significance of ethnoastronomy

All of humankind has a sky, and the night sky remains prominent in many of the world's non-industrialised nations that have so far escaped being brightly lit at night. Since the sky forms a prominent part of the perceived environment, indigenous perceptions of the sky form an integral part of indigenous cosmologies.

Generations of ethnographers, and more recently a few specialist ethnoastronomers, have investigated other 'astronomies'—other perceptions concerning the sky—and their social implications. Some books offer detailed case studies in ethnoastronomy (e.g. [17, 18, 19]) and there are a few compilations (e.g. [20, 21]), but most of this work is scattered among the broader anthropological or astronomical literature and, more recently (since there is no clear dividing line between 'pure' ethnography and ethnohistorical, historical, and even archaeological investigations of indigenous astronomies), in the broader literature on 'cultural astronomy' (e.g. [12, 22]).

The authors, even those who are not trained anthropologists, are generally well aware of the limitations of their attempts to understand and interpret "the [mental] constructions that living people have devised based on their observations and interpretations of the celestial sphere shared by all of us" [23]. Some recurring themes are more obvious and easy to pursue than others. Thus the connections between indigenous astronomical observations and seasonal practices and calendrics are widely recognised and investigated: "Non-Western astronomy is frequently based on a view of the world that is cyclical and organic, intricately linking people and nature in such a manner that there is no essential division between the two. … Because of this, … the practice of astronomy is a central part of socio-cultural experience for non-Western peoples" [24].

From a Western perspective it is tempting to see the immutable cycles of the sky as a passive backdrop—a convenient 'reference system' that helps regulate seasonal and subsistence activities. In an indigenous understanding, however, asterisms may well be active participants in the lifeworld. To the Colombian Barasana, for example, the appearance of the 'caterpillar jaguar' constellation [25] does not just happen to coincide with the appearance of earthly caterpillars (an important food source), so that (as it would seem to us) the celestial event acts as a convenient marker for the terrestrial one; in their perception the Father of Caterpillars, by rising higher and higher in the sky, is directly responsible for the increasing numbers of earthly caterpillars. The stars may also reflect (and affect) society and other aspects of the world in various ways. In ancient Hawaii, for example, social hierarchies were played out in the sky: traditional genealogies identify chiefly, priestly and commoner stars [26].

Ethnoastronomy, and 'cultural astronomy' in general, has already provided plenty of evidence concerning other world-views—fundamentally different ways of conceptualising the cosmos—by focusing on (what we see as) an immutable visual resource common to all of humanity. From a Western perspective, cultural astronomy may certainly be deemed worthy of attention as a sub-field within anthropology giving its own particular insights into human cultural diversity on a cognitive level. But it also appeals to specialist ethno- and archaeoastronomers, perhaps because it gives Western investigators a better understanding of non-Western concepts and practices that, while differing absolutely from the Western model, are, in the broadest sense, 'scientific'.

But how do we bridge the divide—how do we, as Western scientists and ethnographers, progress from merely describing indigenous astronomy to our own satisfaction and try to proceed to a level of discourse that might affect our progress in science as well as our anthropology?

### 3.2 Ethnoastronomical fieldwork: techniques and limitations

While there is no correct, 'objective' way of undertaking ethnoastronomical fieldwork, such research has to be guided by certain general principles that apply equally to ethnographic research in general [27]. Ethnographers have long had the goal of trying "to grasp the native's point of view, his relation to life, *his* vision of *his* world" [28], and so attempt to establish as full as possible a context for communication by learning, in as may ways as possible, to see things as their informants see them. This means living within the community, participating in as wide a range of activities as possible, learning the language and the concepts embedded in it, avoiding taboos, and so on, while all the time not being judgemental and being willing to 'give' as well as 'take'. At the same time it is essential to be aware of the critical distinction between the 'emic' perspective of the insider/participant/actor and the 'etic' perspective of the outsider/observer. As Aveni puts it, the process of doing ethnoastronomical fieldwork can be characterised by the need "to reconcile emic evidence rooted in contemporary native religious belief systems regarding their ancestors with that acquired in the field by the etic approach of the outside investigator" [29].

There are, of course, all manner of potential pitfalls awaiting the inexperienced ethnographer on a basic level. For example, it is critical to ask things that make sense in 'their' framework so that the answer received at least partially fits the intended question. It is important to be aware that the informant may avoid the sacred truth, which cannot be communicated, or to try to please the outsider by giving them the answers they are perceived to want. It is important to beware what one might call the 'warrior' syndrome (encountered by the current author in Polynesia) where the informant, if they do not know the answer to a question, must always "make up an answer on the spot and stick to it". And so on.

The various basic techniques that can be employed include continual efforts (without offending) to reinforce stories in order to so see if they are repeated [27]. At the same time, one may have to counter the tendency to seek more widespread frameworks of knowledge than actually exist. Traditional knowledge of the skies can be very localised, and even personal, as among the Inuit, for whom knowledge of celestial phenomena is "in varying degrees, specific to communities, families … When imparting information, elders frequently made it plain that they were speaking for themselves, that their opinions were not necessarily correct in any absolute sense, and that other elders might, and in probability did, have different views" [30]. If indigenous cosmologies are moulded by a process of cultural consensus [14], then it is always possible that very little consensus actually exists.

It is also important to realise that while some frameworks of cosmological thought have persisted for generations, and even millennia (an obvious case being imperial China [31]), indigenous world-views (especially among smaller, less socially complex human societies) are generally dynamic and have been continually adapted since time immemorial in the light of new experience. This includes contact with modern views: thus the anthropologist Jane Young recounts how Zuni informants 'adopted' the (thoroughly discredited but still popular) '1054 supernova' explanation of a collection of pictographs in Chaco Canyon, New Mexico [21].

As already discussed, important new insights can be imparted by indigenous scholars steeped in the their own cultural tradition. An excellent example, which encompasses a living religious tradition, ethnohistory, history, and even archaeology is the work of the Hindu scholar Rana P.B. Singh on the cosmic symbolism embodied in the sacred cities and landscapes of India [33]. Another approach is for a Western scholar to work alongside an indigenous one. On the other hand, the mere fact that a scholar is himself/herself indigenous should not, and does not necessarily, place their ideas above criticism from other scholars, including Western ones. An example is the 'astronomical and directional register' explanation of a heiau (temple enclosure) in Hawaii [34], which has been widely criticised despite one of the authors being a very well respected Native Hawaiian scholar with considerable knowledge of indigenous astronomy. This predicament may seem more conceivable in archaeoastronomy, where an indigenous scholar (just as much as an outsider) is striving to interpret an archaeological record well separated from any living tradition, but could arise equally within ethnoastronomy.

### 3.3 Drawing the line: new-ageism illustrates the challenge

We mentioned earlier the conference speaker who suddenly revealed himself to be share indigenous beliefs, and the need to treat these views with scholarly respect rather than scientific disdain. But where should the line be drawn?

The first issue to be addressed in this regard is contemporary religious belief. Religion does not have to be incompatible with science, as a number of prominent scientists have argued (e.g. [35, 36]). A major issue, of course, is to reconcile the scientific premise of an objective universe and the religious premise of the power of conscience to influence it; yet anyone, whatever their own beliefs, can easily acknowledge that this is not an issue for other worldviews where human agency and intervention is a self-evident feature of the 'lifeworld'. It is easy to argue that the best people to explore (say) the Hindu historical tradition are likely to be practising Hindus steeped in the tradition.

What, though, of 'alternative' religions? Druidry, for example, has just been recognised under UK charity law as a bona fide religion giving it equal status to major world religions such as Christianity, Islam and Judaism [37]. Of course, this has overtones in relation to archaeoastronomy, in that Druids famously converge on Stonehenge at the summer solstice, despite the fact that, as the current author is fond of saying, "modern Druids have as much to do with ancient Druids as ancient Druids have to do with Stonehenge, which is nothing". The key issue here, however, is whether (modern) Druids could also be 'scientists' (in the broad sense): in so far as Druidry is based on the belief that nature is sacred, and one aspect of its popularity is that it encourages people to live with respect for the environment, this issue night not be as contentious as it might otherwise seem. It is perhaps not out of the question that a practicing Druid could conceivably become a respected professional archaeologist.

The degree of contention becomes greater when we move on to astrology. Can practicing astrologers be taken seriously as modern academics? Consider the following example. The most plausible explanations of the 'Star of Bethlehem' were not achieved by the looking for literal conjunctions of bright objects in the night skies around 6 BC but by adopting a history of astrology approach, analysing the astrological interpretation of the configuration of the sky within the mindset of the Magi, astrologers from the east [38]. This work was actually done by an astronomer, but what would have been our reaction if it had been carried out by a practising astrologer, many of whom have a very active interest in the history of their subject? The answer from most Western astronomers would be to question how we could possibly take seriously the academic credentials of someone who seriously believes that the relative positions of the heavenly bodies as seen from the surface of the earth directly influence individual human lives. Yet is it right that, as Nick Campion has observed, "a western astronomer who may be angry at, or contemptuous of, western astrology, would probably not dare express the same sentiments regarding, say, Native American cosmology or sub-Saharan celestial divination" [39]? It must at least be acknowledged, in the light of the foregoing discussions, that this is not a closed issue.

The final step in this progression through increasingly unpalatable 'alternatives' is surely the various 'new age' ideas that have been perpetrated within Western society itself during the last half-century or so. An early example is the 'ley-line' phenomenon that achieved extraordinary popularity in Britain during the 1960s and 1970s [40]. This popularity derived, at least in part, from the fact that 'ley hunting' satisfied people's desire to be involved in making first-hand discoveries while feeling left behind by 'real' science, for which they were unqualified. As a result, although presented as an alternative to 'closed-minded' science, it led ironically to attempts to emulate science by adopting pseudo-scientific methods, often in extraordinary detail [41].

The reason for mentioning this example here is that similar 'new age' ideas have begun in more recent years to be presented as indigenous beliefs. In the 1970s, archaeological excavations of prehistoric megaliths in Britain were sometimes disrupted by people concerned about the disturbance to 'lines of force', but how should we react when, as happened to a colleague of the author's, a survey of a prehistoric mound site in North America is challenged by Native Americans on similar grounds? The issue, of course, is to decide whether we are prepared to accept that a 'new age' concept such as 'lines of force' has 'genuinely' been incorporated into an indigenous world-view that may well have included a concept broadly akin to 'earth force' in the first place, or whether the 'indigenous' label is being cynically exploited. There is no clear dividing line between these two possibilities.

In an ideal world, we might seek to make a clear distinction between 'genuine' cultural world-views (that do not have access to a range of evidence from modern science and make sense of what is observed and experienced in a logical way, albeit using a logic different from Western logic) from 'new-age' belief systems (that are fully cognisant of, but choose to challenge and dismiss, Western science). In the real world it is not that easy.

We have argued for open-mindedness and tolerance, but in practice how far can and should that tolerance be pushed? Once more this establishes education and communication as the central issue—but one that must be understood from the outset not as a one-way but as a two-way process.

### 3.4 Ways forward

The preceding discussion has posed a series of questions and no easy answers. It is clear that ethnoastronomers need to move on from Western description and analysis of non-Western concepts. If 'cultural astronomy' is to be an all-inclusive activity rather than just a Western one then the whole process of scholarship in the field has to be one that is shared on a basis of mutual cooperation and respect between different cultural viewpoints.

Campion [39] has discussed a number of issues related to those discussed here and has proposed some radical answers, which he has started to try to implement in the context of cultural astronomy courses being taught at the University of Wales, Lampeter. To paraphrase, some of the main components of his approach are the following.

*Neutrality*. Even contemporary ideas that seem open to ridicule (astrology, new-ageism, creationism, UFOs, etc) should be examined and understood as fully as possible, as sociological phenomena, along with one's attempts to examine and understand 'other' world-views. This means (for a Western participant, as well as for others) looking dispassionately at one's own culture and tackling issues that may be politically highly charged; but it is the only way to assure methodological neutrality.

*Participation*. This not only covers 'genuinely experiencing' alien beliefs and practices in the sense of an ethnographer doing fieldwork, but also encompasses the phenomenological objective of 'experiencing' at first hand sky phenomena such as a truly dark night sky, and light-and-shadow hierophanies at key archaeoastronomical sites. (Campion also raises the question of whether it should extend to working with practising astrologers.)

*Reflexivity*. This involves examining one's one concepts and beliefs, so as to be able to 'deconstruct' some aspects of one's own conceptual framework in order better to understand others'. It should also help one to 'listen' more clearly and less judgmentally when communicating 'across the divide'.

If cultural astronomy succeeds in achieving some of the goals identified in the preceding discussion, then we may be on the way to developing a truly global 'anthropology of astronomy' [42]. A necessary part of this project is for contemporary astronomers to reflect on their own beliefs, and this may also go some way to answering the question of whether cultural astronomy has any relevance to progress in modern astronomy. What happens, then, if we try to view modern cosmology in similar terms to the indigenous cosmologies we have been discussing so far? We will briefly consider this issue in the following section.

## 4. Indigenous cosmologies and modern cosmology

One question asked at the Special Session was whether people will look back on the twentieth century as a time when the Western world-view underwent fundamental change. In the light of the IYA 2009 being the 400th anniversary of the astronomical use of the telescope, the question could be extended to cover the previous four centuries. While it is possible to draw attention to major 'scientific revolutions' such as those that successively removed the earth, the solar system, and our galaxy from a special place in the Universe, or to the discovery of the big bang and the Hubble expansion, one could also argue that they are all built upon the fundamental tenets of Western scientific rationality and that these have defined a world-view so successful that it has sustained all the huge cosmological advances of the past 400 years and more and is set to do so indefinitely. A different line of argument might identify quantum mechanics as the one development that has fundamentally changed our world-view, on the grounds that the collapse of the wave function following an observation appears to demolish the Cartesian duality between subject (the observer) and object (the 'external' universe).

However, to pursue any of these lines of argument in isolation is to ignore the social dimension that is inherent in all indigenous cosmologies. In this sense, Primack and Adams [43] may be nearer the mark when they observe that at no time in human history have the 'big questions' about the nature of the universe been so removed from everyday people and everyday lives. As they point out, this separation has resulted in a quite unprecedented level of 'cultural indifference' to the unimaginable wonders about the universe that science is revealing. Put another way, we can say that within modern Western culture there is an unprecedented level of indifference to modern cosmology compared with what was true in past cultures—for whom the skies impacted much more prominently on their everyday existence—or indeed compared to many indigenous populations today.

It is wrong, however, to conclude that many of today's population have no cosmology at all. Modern scientific cosmology may have become irrelevant for people at large, but 'universal concerns' continue to feature prominently in world religions and cult beliefs, as well as in various 'alterative' beliefs such as astrology and new-ageism. In many senses, these can be regarded as elements in the 'indigenous cosmologies' of a great many people living in Western societies today. Then again, science and religious belief may seem incompatible to many, but (as we have seen) even in today's world many people find a way to construct a more complex philosophical underpinning—in a sense, their own world-view—that sustains both. All this suggests that instead of envisaging the scientific world-view as 'the' Western world-view, we should actually entertain a more complex picture of Western society where a variety of world-views co-exist. This has the advantage of laying the way for a coherent approach towards 'alternative' beliefs that resolves the inconsistencies identified by Campion. However, in doing so it could also be taken to suggest a level of tolerance towards 'alternative' philosophies, raising 'new-ageism' to the level of 'genuine' indigenous beliefs, that many people (this author included) would find quite unacceptable.

Primack and Adams' solution is to formulate a 'new cosmology' to fill the 'vacuum' that they perceive. It is one where human life can still be seen as central in various ways against the backdrop of the universe as described by modern astronomy. The aim is to restore people's engagement with cosmic problems (and, at the same time, global problems). Laudable as this goal may be—'progressing' our cosmology in a sense—achieving it may be a good deal more difficult even than Primack and Adams envisage, since (as we have argued) there is really no 'vacuum' in the first place, but rather a confusion of 'alternative' world-views. Directly challenging them is surely the wrong approach. In fact, we can paraphrase a remark that we made back in §2.1: when one is dealing face-to-face with contemporary 'alternative' world-views, any attempt (explicit or implicit) to place modern science on a pedestal will almost inevitably be taken as condescending and patronising.

In fact, the discussions of §§2 and 3 raise a set of issues that are equally relevant in approaching the various 'divides' within Western society itself. Attempts such as Primack and Abrams' to seek reconciliation may be crucial, yet their own arguments are full of phrases such as 'this is how it *really* works', references to 'pre-scientific cultures', and so on. Open-mindedness and tolerance can be extremely difficult to achieve when the superiority of the scientific method seems pre-evident, and it can be almost impossible to avoid the sort of terminology that betrays this implicit assumption. Yet if we wish to engender a more universal 'respect' for modern cosmology, bearing in mind that a two-way process is involved as for 'other' indigenous cosmologies, the only way may be to strive better to understand, and hence better to respect, on some level, at least some of the most prevalent alternative world-views that co-exist with it. Indeed, this may be the only way to reach effectively into the conceptual melee that characterises modern Western society, and within this melee dwell many of the political and private supporters and sponsors without whom developments in modern science would be impossible.

A rather different situation is evident in some 'Westernised' societies where a strong sense of indigenous cultural identity still persists. In Japan, for example, the perception that Japanese native beliefs and practices are somehow inferior to those of imported cultures (and, by implication, that indigenous astronomies are somehow inferior to modern scientific astronomy) has led to a lack of cultural self-esteem [44]. By contributing to a general sense of indifference, a lack of cultural esteem not only fails to act in the interests of scientific and technological progress but arguably hinders it. Cultural esteem relates to cultural identity, which operates at a variety of levels and is an important issue in all human societies. A high level of cultural esteem is clearly predicated on the view that local indigenous perspectives are *not* inferior, but by reducing indifference to cultural progress it may also be a key factor in supporting the coexistence of the view that progress in global science is also to be supported and encouraged. It can be relevant in resolving potential conflicts such as indigenous Hawaiian concerns about the Mauna Kea Observatory. Additionally, it can both encourage and be further enhanced by the development of projects that 'use' modern science for cultural ends, such as one of the outcomes of the TOPS workshops organised by the Institute for Astronomy in Hawaii in the early 2000s (http://www.ifa.hawaii.edu/tops/) which encouraged community colleges to teach local students the techniques and methods of archaeoastronomy.

Astronomical heritage is also of considerable relevance in this context. The focus here is not upon historical observatories but upon sites that, through their layout and astronomical alignments, provide valuable insights concerning ancient cosmologies or bear witness to more recent indigenous ones. Cultural identities operate at regional, national and local levels, and heritage in general, both tangible and intangible, reflects (or is taken to reflect) cultural histories and hence helps to reinforce modern cultural identities and esteem. One does not have to see such sites as icons of a Westernised view of global developments in astronomy (e.g. Stonehenge or Plains Indian 'medicine wheels' viewed as 'observatories') in order to accept that they represent local developments that are important on a global level in demonstrating the richness and diversity in ways in which human communities have understood and interpreted the sky, and hence can serve to generate enthusiasm and support for progress in modern astronomy. This is perhaps best summarised in the title of the concluding section of Renshaw and Ihara's overview of developments in Japanese star lore and mythology [44]: "Hopes for the future built on a rich but hidden past".

We cannot conclude this discussion without mentioning another type of heritage, which is of fundamental relevance: the natural heritage of the dark night sky. A substantial proportion of the world's population (especially in Western industrialised countries) may never in their lifetimes see (or even get the opportunity to see) a truly dark night sky; light pollution has served to remove them from 'their' sky as a source of inspiration. This not only makes it more difficult for them to appreciate how important the sky is to the world's indigenous peoples but also contributes to the feeling that modern astronomy has no relevance for them. The preservation of some of the world's darkest night skies is a common cause that unites many indigenous peoples, who wish to preserve their own pristine skies, and professional astronomers who are concerned to preserve the world's largest ground-based optical observatories [45].

These rather diverse arguments can perhaps be best summed up by saying that an approach that deals integrally with traditional or indigenous astronomies alongside modern science can help strengthen cultural identities while at the same time facilitating scientific progress. The extent to which 'alternative' world-views within Western society should be treated in a similar way to 'genuine' indigenous cosmologies will continue to generate debate and contention, but we do at least have to acknowledge the existence of several, and possibly a multitude, of disparate world-views within Western society and that in order to engender wider appreciation and support for progress in modern astronomy and cosmology we have to understand, achieve a level of respect for, and develop serious communications with the proponents of at least some of the major ones.

Some might worry, conversely, that introducing ideas of modern science in indigenous contexts risks distorting or even destroying indigenous systems of thought, with a consequent negative effect upon human cultural diversity. While it surely imperative to try to record aspects of indigenous knowledge for the posterity of humankind, it is also arguable that this is essentially a heritage issue and should not stand in the way of progress (in the sense of introducing modern ideas into indigenous systems of knowledge). It is worth recalling that indigenous world-views are dynamic and have been continually adapted since time immemorial in the light of new evidence. Valuing indigenous systems of thought does not mean trying to 'freeze' them—indeed, the condescension that would be implicit in doing so implies that the very opposite is in fact the case—and a corollary of this is that we need feel no guilt about outreaching 'our' science, provided this is undertaken in a spirit and framework of mutual understanding and respect.

## 5. Conclusion

What, from a 'Western' perspective, may be a genuine desire to communicate and share our knowledge and excitement about the universe and to open the way for a wider range of the world's population to become actively involved may be perceived outside the 'Western' world as innately wrong-headed and—worse—patronising, detrimental, and even colonial. In order to overcome this problem we need firstly to realise that it is really there: that a barrier exists. The first step in approaching it, as we have suggested, is to try to counter negative of indigenous perceptions of 'Western' astronomy, and in particular to recognise the assumption (either explicit or implicit) that our scientific viewpoint is superior, rather than just different, as not only counterproductive but potentially offensive.

In a modern world where widespread public perceptions include many that are unscientific in the broadest sense of the term, there are actually a range of positive benefits in terms of public support for progress in scientific astronomy (and all that follows from that) in recognizing, acknowledging, and respecting 'genuine' cultural world-views whose goals—in common with those of modern science—are to make sense of the cosmos within which we live. While mutual contact and education is clearly helpful, the necessary level of reconciliation can only be achieved through more fundamental attempts by modern astronomers to understand, and ultimately to respect, the non-western frameworks of thought that give rise to other cultural perspectives—as well as the heritage associated with them.

*Indigenous Astronomies* Clive Ruggles